\documentclass[usenatbib]{mn2e}

\usepackage{graphicx}
\usepackage{times}

\begin{document}

\title[ Accretion Disc Age and Stellar Age]{The Relationship between Accretion Disc Age and Stellar Age and its Consequences for Proto-Stellar Discs}
\author[M G Jones, J E Pringle \& R D Alexander]
{Michael G. Jones$^1$, J.E.Pringle$^2$ and R.D.Alexander$^3$\\$^1$ Fitzwilliam College, Cambridge\\$^2$ Institute of Astronomy, Madingley Road, Cambridge CB3 0HA, UK \\$^3$ Department of Physics \& Astronomy, University of Leicester, Leicester LE1 7RH, UK}
\date{22/07/11}
\maketitle

\begin{abstract}
We show that for young stars which are still accreting and for which measurements of stellar age, $t_\ast$, disc mass $M_{\rm disc}$ and accretion rate $\dot{M}$ are available,  nominal disc age $t_{\mathrm{disc}} = M_{\mathrm{disc}}/\dot{M}$ is approximately equal to the stellar age $t_\ast$, at least within the considerable observational scatter. We then consider theoretical models of proto-stellar discs through analytic and numerical models.  A variety of viscosity prescriptions including empirical power laws, magnetohydrodynamic turbulence and gravitational instability were considered within models describing the disc phenomena of dead zones, photoevaporation and planet formation.  These models are generally poor fits to the observational data, showing values of $t_{\rm disc}$ which are too high by factors of 3 -- 10. We then ask whether a systematic error in the measurement of one of the observational quantities might provide a reasonable explanation for this discrepancy. We show that for the observed systems only disc mass shows a systematic dependence on the value of $t_{\rm disc}/t_\ast$ and we note that a systematic underestimate of the value of disc mass by a factor of around 3 -- 5, would account for the discrepancy between theory and observations. 
\\
\\
\end {abstract}

\section{Introduction}
Proto-stellar discs form during the star formation process.  Much of the material falling inwards during the final collapse phase has non-zero angular momentum about the newly formed proto-star.  This angular momentum must be removed in order for the material to fall further inwards, and this is the basis of an accretion disc (\cite{Williams2011, Armitage2011} and references within).

The inward falling material flattens into a disc and then viscous processes, probably driven by a combination of magnetohydrodynamic turbulence and gravitational instabilities, redistribute the angular momentum and allow accretion on to the central star \citep{Pringle1981}.  Within a few tens of Myr the gas in the disc has either accreted on to the star or been blown away by stellar radiation, while dust grains have aggregated, leaving behind a debris disc and potentially a planetary system.

Having a detailed model of these systems is vital to understanding both the late stages of star formation and planet formation processes.  At present discs are poorly understood and many of their parameters are only weakly constrained.  These loose constraints stem from the difficulties involved in observing systems which subtend just $\sim\!1\arcsec$ on the sky and for which detection of their main constituent, molecular hydrogen, is not currently feasible \citep{Andrews2009}.  Despite this, such discs are often observed around T Tauri stars through the IR excess they produce and emission from their dust in the millimetre and sub-millimetre.  As the temperature of the disc changes in the radial direction away from the central star (as does the disc structure), the spectral energy distribution can be used to probe the disc structure without it needing to be spatially resolved \citep{Beckwith1990}.  However, large uncertainties remain in many of the disc properties, principally due to the assumptions required in order to proceed in this manner. 

In general the equations describing the evolution of accretion discs are non-linear and require numerical treatments, but if simplifying assumptions about the form of the viscosity are made, then analytic solutions are possible \citep{vonWeizsacker1948,Lust1952,Lynden-Bell1974}.  These solutions give the total disc mass decaying as $M_{\rm disc} \propto t^{-\sigma}$, where $\sigma>0$.  Differentiating this leads to the accretion rate onto the central star behaving as $\dot{M} \propto t^{-(\sigma+1)}$.  For this reason, as well as on purely dimensional grounds,  one can define an age indicator, namely the disc age $t_{\rm disc}$ as

\begin{equation}
t_{\mathrm{disc}} = \frac{M_{\mathrm{disc}}}{\dot{M}}.
\label{eqn:discage}
\end{equation}
Disc age is the characteristic accretion timescale of the disc at any given time.  For simple viscous disc models disc age is proportional to the age of the accretion disc with a proportionality constant of order unity.  Therefore we regard disc age as a proxy for the age of the system, which in principle should be related to the age of the central star, $t_\ast$.  Disc age is measurable as both $M_{\rm disc}$ \& $\dot{M}$ are observable quantities, and can be measured from independent observations (see \textsection 2).

In Section 2 we summarise the techniques involved in making the observational measurements and display the observational data in the $t_\ast - t_{\rm disc}$ plane.  In Section 3 we then investigate various simple theoretical models of accretion disc evolution and show that these models tend to produce values of $t_{\rm disc}$ that are too high by factors of 3 -- 10. We show that taking account of various complications such as photoevaporation and planet formation do not account for this discrepancy. In Section 4 we discuss these results and investigate the possibility that the discrepancy might be accounted for by a systematic error in one of the observed quantities, concluding that a possible solution is that the disc masses are systematically underestimated by a factor in the range 3 -- 5.

\section{Observational Data}

\begin{figure}
\centering
\includegraphics[width=1\columnwidth]{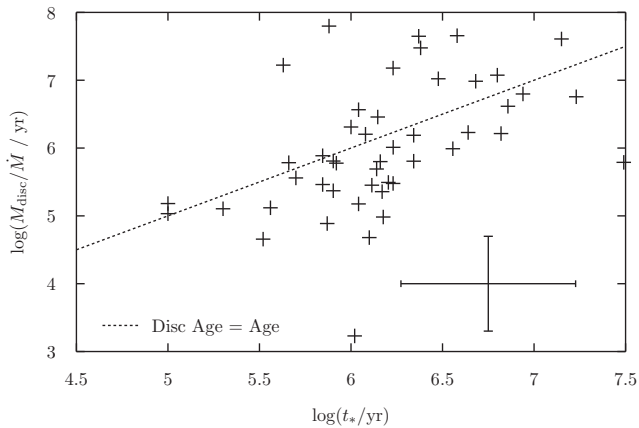}
\caption{Observations of young stellar objects with accretion discs.  Nominal disc age, $t_{\rm disc}$, (y-axis) plotted against stellar age, $t_\ast$, (x-axis) on a logarithmic scale.  The line $t_{\rm disc} = t_\ast$ is included for comparison. This line, $t_{\rm disc} = t_\ast$, is a reasonable fit to the data, although there is a high degree of scatter, consistent with observational uncertainties.  The horizontal error bar is a factor of 3 (the error estimate of stellar ages).  The vertical error bar is a factor of 5 from the combination of the error estimates of accretion rate and disc mass (factors of 3 and 4 respectively).  The detection limits on accretion rate and disc mass are approximately $10^{-10}$ $M_\odot$yr$^{-1}$ \& $0.001$ $M_\odot$ respectively.  This may lead to preferential exclusion of data points in the upper right of the plot, as disc mass generally decreases with age.  The minimum stellar age cut off is $\sim\!10^5$ Myr.}
\label{fig:obsplot}
\end{figure}

\begin{table*}
\begin{minipage}{110mm}
\caption{Observational data on young stellar objects with accretion discs}
\label{tab:obs}
\begin{tabular}{@{}lccccr}
Object & $\log(t_{*}/\mathrm{yr})$ & $M_{\mathrm{disc}}/M_{\odot}$ & $\log(\dot{M}/M_{\odot}\mathrm{yr}^{-1})$ & $\log(t_{\mathrm{disc}}/\mathrm{yr})$ & Source\\
\hline
AA Tau&6.38&0.0825&-8.56&7.48&R\\
CI Tau&6.23&0.0115&-7.95&6.01&R\\
CW Tau&6.82&0.0043&-8.58&6.21&R\\
CX Tau&5.90&0.0016&-8.18&5.37&R\\
CY Tau&6.15&0.0165&-8.24&6.46&R\\
DE Tau&5.30&0.0026&-7.69&5.10&R\\
DL Tau&6.34&0.0140&-7.66&5.81&R\\
DM Tau&6.56&0.0100&-7.99&5.99&R\\
DN Tau&6.04&0.0104&-8.55&6.56&R\\
DO Tau&6.11&0.0225&-7.10&5.45&R\\
DR Tau&6.18&0.0073&-7.12&4.98&R\\
DS Tau&7.23&0.0570&-8.00&6.76&R\\
FM Tau&6.80&0.0880&-8.13&7.07&R\\
FZ Tau&5.90&0.0466&-7.14&5.81&R\\
GM Aur&6.86&0.0150&-8.44&6.62&R\\
GO Tau&6.68&0.0070&-9.14&6.99&R\\
HO Tau&6.94&0.0047&-9.13&6.80&R\\
IQ Tau&6.08&0.0092&-8.24&6.20&R\\
RY Tau&6.04&0.0140&-7.03&5.18&R\\
SU Aur&6.34&0.0466&-7.52&6.19&R\\
UZ Tau E&5.00&0.0103&-7.02&5.03&R\\
AS 205 (A)&5.00&0.1100&-6.14&5.18&A7\\
DH Tau (A)&5.92&0.0030&-8.30&5.78&A7\\
Elias 24&5.66&0.1300&-6.67&5.78&A7\\
SR 24 (S)&6.64&0.1200&-7.15&6.23&A7\\
AS 205&5.70&0.0290&-7.10&5.56&A9\\
AS 209&6.20&0.0280&-7.05&5.49&A9\\
DoAr 25&6.58&0.1360&-8.52&7.66&A9\\
GSS 39&6.00&0.1430&-7.15&6.31&A9\\
VSSG 1&5.85&0.0290&-7.00&5.46&A9\\
WaOph 6&5.85&0.0770&-7.00&5.89&A9\\
WSB 60&6.48&0.0210&-8.70&7.02&A9\\
DD Tau&6.14&0.0020&-8.39&5.69&H\\
DK Tau&5.56&0.0050&-7.42&5.12&H\\
DP Tau&6.17&0.0030&-7.88&5.36&H\\
DQ Tau&5.88&0.0250&-9.40&7.80&H\\
FV Tau&6.02&0.0010&-6.23&3.23&H\\
FY Tau&5.87&0.0030&-7.41&4.89&H\\
GG Tau&5.63&0.2900&-7.76&7.22&H\\
IP Tau&6.23&0.0120&-9.10&7.18&H\\
LkCa 15&6.37&0.0600&-8.87&7.65&H\\
UY Aur&5.52&0.0030&-7.18&4.66&H\\
BP Tau&6.16&0.0182&-7.54&5.80&V\\
DF Tau&3.59&0.0004&-6.75&3.35&V\\
DG Tau&6.10&0.0240&-6.30&4.68&V\\
Haro 6-28&6.23&0.0006&-8.70&5.48&V\\
HL Tau&7.15&0.0600&-8.83&7.61&V\\
HN Tau&7.49&0.0008&-8.89&5.79&V\\
\hline
\end{tabular}
Column 1: Object name; Column 2: $\log_{10}$ of stellar age estimate (yr); Column 3: Disc mass estimate ($M_{\odot}$), the dust masses from source R were averaged between $q=2.5\&3$, and multiplied by 100 to give total disc mass; Column 4: $\log_{10}$ of stellar accretion rate ($M_{\odot}\,\mathrm{yr}^{-1}$); Column 5: $\log_{10}$ of disc age ($=M_{\mathrm{disc}}/\dot{M}$) in years; Column 6: Sources R = \cite{Ricci2010}, A7 = \cite{Andrews2007}, A9 = \cite{Andrews2009}, H = \cite{Hartmann1998}, V = Various \citep{White2001,Najita2007,Andrews2005,Muzerolle2003,Gullbring1998}.
\end{minipage}
\end{table*}

In principle the three properties required to compare the disc age $t_{\rm disc}$ with the stellar age $t_\ast$ for a given system (accretion rate, disc mass, and stellar age, taken as the system age) are all measurable.  The requisite observational data, taken from a variety of sources, are presented in Figure \ref{fig:obsplot} and Table \ref{tab:obs}.  In this section we give a brief outline of the procedures behind the measurement of each property. For a more detailed description we refer the reader to the referenced sources.

\subsection{Stellar Age $t_\ast$}
\label{sec:age}
Spectroscopy is used to determine the spectral type and effective temperature of each star.  Combined with the measured luminosity this allows the star to be located in the Hertzsprung-Russell diagram (hereinafter H-R diagram).  From here theoretical pre-main-sequence evolutionary tracks are used to convert the H-R diagram position to a stellar age and mass.

As with all astrophysical sources the light from these systems is reddened by intervening material and this must be corrected for.  However, the extinction correction process may well be confused by the presence of a circum-stellar disc.  Accretion onto the star creates a blue excess and emission from the disc creates a red excess, hence making the spectrum both too blue and too red \citep{Hillenbrand2009}.  In addition starlight may be either reflected off, or extinguished by, a disc, either increasing or decreasing the observed luminosity.

Also, there is uncertainty about which theoretical tracks are the most relevant, and different authors use different tracks, often leading to substantially different age estimates, particularly for the youngest stars.  \cite{Hillenbrand2009} estimates the errors to be approximately a factor of 3 in either direction.  The fractional errors are likely to reduce as the stars become older.

\subsection{Stellar Accretion Rate $\dot{M}$}
If the distance and spectral type are known, stellar radii can be found from luminosities using the Stephan-Boltzmann law.  Material is then assumed to fall from the inner edge of the disc (usually taken as $R_{\rm in} \approx 5 R_\ast$) onto the surface of the star, releasing its gravitational potential energy as accretion luminosity, given by

\begin{equation}
L_{acc}=\frac{GM_{*}\dot{M}}{R_{*}}\left(1-\frac{R_{*}}{R_{\rm{in}}}\right),
\label{eqn:acclum}
\end{equation}
where $M_*$ is the stellar mass derived from evolutionary tracks, $\dot{M}$ is the accretion rate, $R_{*}$ is the stellar radius.

The characteristic temperature of this emission is of the order of 10,000 K, which is greater than the stellar emission temperature at a few thousand K.  The UV excess is measured and taken to be entirely due to the accreting material, giving the accretion rate through the above relation.

Uncertainty in the distance to the sources could be a substantial source of systematic error.  Converting from stellar luminosity to radius introduces one factor of distance, $d$,and converting from observed accretion flux to luminosity adds two more, meaning the final value depends on $d^{3}$.  The other key source of error is the stellar mass derived from evolutionary tracks.  Final errors in the accretion rate were taken to be a factor of 3 \citep{Hartmann1998, Calvet2000}.

\subsection{Disc Mass $M_{\rm disc}$}
\label{sec:obsmass}
Most of the mass of a proto-stellar disc is at large radii and is contained in molecular hydrogen. However this is a symmetric molecule with no permanent electric dipole, making it difficult to detect. Thus the discs are usually detected through IR emission from warm dust in the inner disc, or through mm or sub-mm emission from cool dust in the outer disc.

To estimate the disc mass from the IR fluxes it is necessary to have a simple model of the disc structure in mind. It is usual to assume surface density structure and a temperature distribution of the forms $\Sigma(R) = \Sigma_{0} R^{-p}$ and $T(R) = T_{0} R^{-q}$ respectively.  Then by assuming a form for the dust opacity (usually $\kappa_{\nu}=0.1(\nu/10^{12}\:\rm{Hz})\;\rm{g\:cm^{-2}}$ \citep{Beckwith1990}) and that the dust emits as a black body at each radius, the free parameters $p$, $q$, $T_{0}$ and $\Sigma_{0}$ can be adjusted to achieve the best fit to the observed spectrum.  Since the IR flux comes from warm dust, this gives the total {\em dust} mass in the disc. The total disc mass is assumed to be 100 times this value. 

The primary source of error here is thought to be the assumed dust opacity. Grains are expected to grow and alter the form of the opacity (as well as the gas to dust ratio), but this process is not well understood and a simple form must be used at present.  

Taking the ISM average dust to gas ratio is probably a reasonable value for the initial disc, but this is likely to change as material aggregates, maybe forms planets and gas is preferentially lost in winds.  However, it is thought that the lack of treatment of grain growth will decrease the mass estimates, as larger bodies are neglected, where as using the ISM dust to gas ratio will over estimate the mass, so there should be some degree of cancellation between these two sources of error.  The overall errors in total disc mass were typically estimated to be a factor of 3 or 4 \citep{Andrews2005,Ricci2010}.

\subsection{Comparison of disc age and stellar age}

As mentioned above we present in Figure~\ref{fig:obsplot}  the observational estimated values for both $t_{\rm disc}$ and $t_\ast$, for those stars for which we were able to discover estimates of mass, accretion rate and age in the literature. We also indicate the sizes of the estimated errors in both these quantities. In Figure~\ref{fig:obsplot}  we also plot the line $t_{\rm disc} = t_\ast$. Given the size of the estimated errors this line seems to provide a reasonable fit to the data, although there is considerable scatter.

\section{Theoretical Models \& Results}
\label{sec:discmodels}

In this Section we investigate the extent to which theoretical models of accretion disc evolution agree with the observational estimates in the $t_\ast - t_{\rm disc}$ plane.\footnotemark[1] \footnotetext[1]{For clarity $t$ will be assumed equal to $t_\ast$ henceforth.  Strictly they are not equal as $t$ is the time since the beginning of the simulation, not the birth of the star.  This approximation improves with age.}

For an axisymmetric matter distribution with surface density $\Sigma(R)$ and viscosity $\nu(R)$ in Kelperian orbits around a central point mass, the surface density evolves according to \citep{Pringle1981}:

\begin{equation}
\frac{\partial \Sigma}{\partial t} = \frac{3}{R} \frac{\partial}{\partial R} \left[ R^{1/2} \frac{\partial}{\partial R} \left( \nu \Sigma R^{1/2} \right) \right].
\label{eqn:accdisc}
\end{equation}
In general $\nu$ is likely to be a function of local disc variables such as $\Sigma$ and $R$, making the equation non-linear and requiring solution by numerical means.  There are however special cases where analytic results can be found.

\subsection{Similarity Solutions}
If $\nu$ is taken to be a power law function of radius $R$ alone then the equation for $\Sigma$ is linear and an analytic solution to the initial value problem is possible by means of a Green's function \citep{Lust1952,Lynden-Bell1974, Pringle1981}.  For viscosities only dependent on a power of the radius ($\nu \propto R^{n}$) the solutions which correspond to zero torque at the origin, so that the total angular momentum of the disc is conserved, but accretion takes place at small radius, take on similarity form at late times:

\begin{equation}
\Sigma = \frac{C}{r^{n}} \tau^{-\rho} \exp \left[-\frac{r^{2-n}}{\tau}\right],
\end{equation}
where $r={R}/{R_{0}}$, $\tau={t}/({t_{0}} + 1)$, $\rho$ is a simple algebraic expression in $n$, and $C$, $R_0$ and $t_0$ are constants.

Thus at late times we find
\begin{equation}
M_{\rm disc} \propto t^{- \sigma},
\end{equation}
where
\begin{equation}
\sigma = \frac{1}{2(2-n)}.
\end{equation}
The requirement that $\sigma > 0$ implies that $n < 2$.

In the more general case in which $\nu \propto \Sigma^{m} R^{n}$ with $m>-1$ for viscous stability (Pringle 1981), there are also similarity solutions. They are of  of the form (Pringle 1991)

\begin{equation}
\Sigma = \Sigma_{0} \left(\frac{R}{R_{0}}\right)^{-\alpha} \left(\frac{t}{t_{0}}\right)^{-\beta} \left[1-k\left(\frac{R}{R_{0}}\right)^{\gamma} \left(\frac{t}{t_{0}}\right)^{-\eta} \right]^{\mu},
\end{equation}
where $\Sigma_{0}$, $R_{0}$ and $t_{0}$ are model dependent constants, and $\alpha$, $\beta$, $\gamma$, $\eta$, $\mu$ and $k$ are simple algebraic expressions in $m$ and $n$.

At late times these solutions also give
\begin{equation}
M_{\rm disc} \propto t^{- \sigma},
\label{mtpowerlaw}
\end{equation}
where now
\begin{equation}
\sigma = \frac{1}{5m + 4 - 2n}.
\label{sigmapowerlaw}
\end{equation}
These similarity solutions correspond to all the mass initially being at the origin at time $t=0$, but with a finite amount of angular momentum. Mass is then accreted at the origin, with the disc spreading to conserve angular momentum. Eventually all the mass accretes onto the central object and all the angular momentum is carried away to infinity by an infinitesimal amount of the mass.  

Because for these solutions $M_{\rm{disc}} \propto t^{-\sigma}$, which can be differentiated to give $\dot{M} \propto \sigma t^{-(1+\sigma)}$ the disc age for these solutions is given by 

\begin{equation}
t_{\rm disc} = \frac{M_{\rm{disc}}}{\dot{M}} = \frac{t}{\sigma}.
\end{equation}
This represents a straight line parallel to $t_{\rm disc} = t_\ast$ in the $\log(t_\ast) - \log(t_{\rm disc})$ plane, but translated vertically by a distance $\log (1/\sigma)$.

\begin{table}
\caption{Opacity regimes $(n=1-12)$ for the opacity law approximated as a series of power law segments $\kappa = \kappa_0 \rho^a T^b \; \mathrm{cm^{2}\,g^{-1}}$.  Maximum temperature $T_{\rm max}$ is given for regimes independent of density \citep{Bell1994, Bell1997}. The value of $\sigma$ relevant to the mass decline in the corresponding similarity solutions (Equations~\ref{mtpowerlaw} \&~\ref{sigmapowerlaw}) is also given for those regimes $n$ which correspond to a thermally and viscously stable disc.}
\label{tab:opc}
\begin{tabular}{@{}lccccr}
$n$ & $\kappa_{0}$ & $a$ & $b$ & $\sigma^{-1}$ & $T_{\mathrm{max}}/\mathrm{K}$ \\
\hline
1&$1.0\times10^{-4}$&0&2.1& 6.79 &132\\
2&$3.0\times10^{0}$&0&-0.01& 3.74 &170\\
3&$1.0\times10^{-2}$&0&1.1& 4.79 &377\\
4&$5.0\times10^{4}$&0&-1.5& 3.00&389\\
5&$1.0\times10^{-1}$&0&0.7& 4.33&579\\
6&$2.0\times10^{15}$&0&-5.2& 2.19&681\\
7&$2.0\times10^{-2}$&0&0.8& 4.44&-\\
8&$2.0\times10^{81}$&1&-24& $\ast$&-\\
9&$1.0\times10^{-8}$&2/3&3& 7.43&-\\
10&$1.0\times10^{36}$&1/3&10& $\ast$&-\\
11&$1.5\times10^{20}$&1&-5/2& 2.86&-\\
12&$0.348$&0&0& 3.75&\\
\hline
\end{tabular}
\end{table}

\subsection{Basic Models}
\label{sec:modres}

We first investigate the properties of some simple models for disc evolution.

We solve the equations governing the evolution of surface density using an explicit first order finite difference method on a radial grid of 100 points, logarithmically spaced in radius $R$.  We take the inner disc boundary radius to be $R_{\rm{in}} = 5\;R_{\odot}$ and the outer boundary radius $R_{\rm{out}} = $  1000 AU. The outer boundary is chosen so that is has little or no effect, except at very late times $\log (t/{\rm yr}) > 7.5 $, when in some models the disc touches the outer radius. The boundary conditions were zero torque ($\Sigma(R_{\rm{in}}) = 0$) at the inner boundary, and zero radial velocity at the outer boundary.  We take the stellar mass to be constant throughout ($M_{*}=0.75\;M_{\odot}$), and so ignore the mass of any added disc material. Typically we assume matter is added to the disc at around $R_0 = 10$ AU with the local disc temperature and specific angular momentum.  Initially all the disc mass is contained in a narrow Gaussian of standard deviation 1 AU centred on $R_0$, with an initial total disc mass of $0.2\;M_{\odot}$.  These conditions hold for all of the following simulations unless explicitly stated otherwise.

\subsubsection{Power-law viscosity}

\begin{figure}
\centering
\includegraphics[width=1\columnwidth]{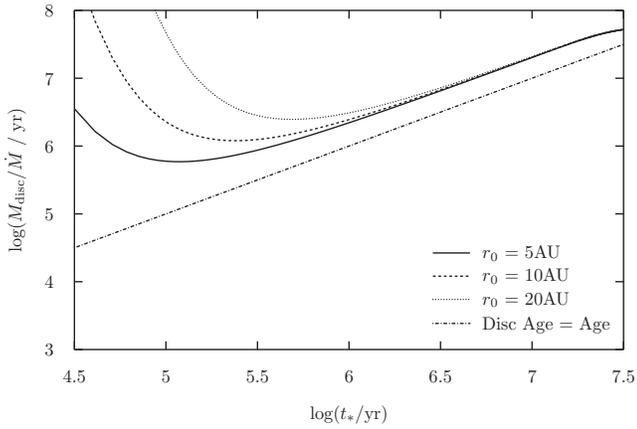}
\caption{Increasing the radius of the centre of the initial mass distribution has a knock on effect on the viscous timescale.  With $\nu \propto R$ the viscous timescale is proportional to $R$, so as the initial disc radius $R_{0}$ is doubled in each consecutive model they should be initially separated by approximately log 2 along the horizontal axis, but still tend towards the same similarity solution at late times.}
\label{fig:r0vary}
\end{figure}

We first consider a simple viscosity of the form
\begin{equation}
\nu = \nu_0 (R/R_0),
\end{equation}
with $\nu_0$ and $R_0$ constants. This viscosity is implied by the common assumption that the disc surface density is of the form $\Sigma \propto R^{-1}$. The loci of the solutions in the $t_\ast-t_{\rm disc}$ plane are given in Figure~\ref{fig:r0vary} where we have also illustrated the effect of varying the initial disc radius $R_0$, for given $\nu_0$. This is, of course, equivalent to varying the size of the viscosity $\nu_0$ for fixed initial disc radius. At late times the solutions approach the similarity solutions, and so $t_{\rm disc}/t_\ast  \rightarrow 1/\sigma = 2$, since here $n=1$. Note, however, that the approach to these solutions is from above, since at early times the accretion rate is lower for a given disc age, since the disc takes time to fully extend to small radii.

\subsubsection{$\alpha$ viscosity}
\label{sec:basicmod}

A more physically motivated approach to the viscosity is to assume
\begin{equation}
\nu = \alpha \frac{c_{\rm{s}}^{2}}{\Omega}, 
\end{equation}
originally from \cite{Shakura1973}, where $c_{\rm{s}}^{2} = \mathcal{R}T/\mu$ is the sound speed squared, $\Omega(R)$ is the orbital angular velocity, $\mathcal{R}$ is the gas constant and $\mu$ is the mean molecular weight, here taken to be 2.3 (appropriate for a molecular disc). The quantity $\alpha$ is the standard dimensionless measure of the viscosity.

In order to implement the $\alpha$ prescription a local model of the disc structure is required. We take the usual one-zone model, which assumes thermal and hydrostatic equilibrium in the local disc annulus. The heating rate (per unit area, for each side of the disc) in the local annulus due to viscous dissipation is

\begin{equation}
Q_{+}(R) = \frac{9}{8} \nu(R) \Sigma(R) \Omega(R)^{2}
\end{equation}
and this is set equal to the local cooling rate due to blackbody emission

\begin{equation}
Q_{-}(R) = \sigma_{\rm SB} T_{\rm{e}}(R)^{4},
\end{equation}
where here $\sigma_{\rm SB}$ is the Stefan-Boltzmann constant and $T_{\rm{e}}$ is the effective surface temperature of the disc.

To specify the surface temperature $T_{\rm{e}}$, assumptions regarding the vertical structure must be made.  We adopt the simplistic vertically averaged approach from \cite{Armitage2001}, giving

\begin{equation}
T_{\rm{e}}^{4} = \frac{8}{3\tau} T_{\rm{c}}^{4},
\end{equation}
where $T_{\rm c}$ is the temperature on the central disc plane.

This, together with hydrostatic equilibrium perpendicular to the disc plane, which gives the disc scale-height
\begin{equation}
H = \frac{c_{\rm s}}{\Omega},
\end{equation}
and the assumption that disc density (on the central plane) is related to surface density by $\rho_c = \Sigma/2H$, form a set of equations which specify the disc structure. We note that then the optical depth $\tau$ is given by
\begin{equation}
\tau = \frac{1}{2} \Sigma \kappa(\rho_{\rm c}, T_{\rm c}).
\end{equation}
Thermal equilibrium is a reasonable assumption if the thermal timescale is much smaller than the viscous timescale \citep{Pringle1986}.  Preliminary tests with a more detailed thermal energy equation showed this approximation was reasonable, and so it was adopted as it was less computationally demanding.

\subsubsection{Single Opacity}

\begin{figure}
\centering
\includegraphics[width=1\columnwidth]{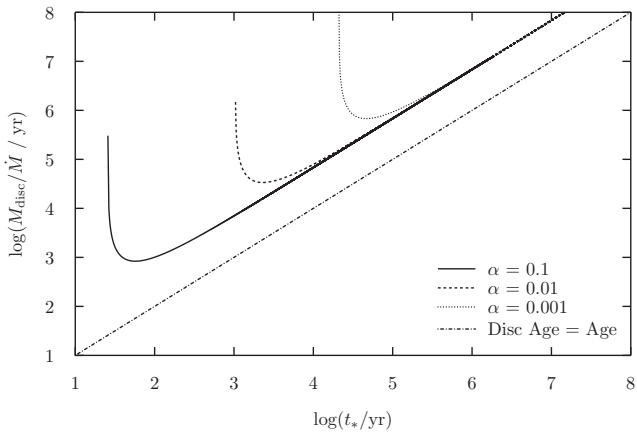}
\caption{Varying the alpha parameter changes the strength of the viscosity, altering the viscous timescale, and hence the time taken to converge to the appropriate similarity solution, which is the same regardless of the value of alpha.}
\label{fig:alphavary}
\end{figure}

We first computed disc evolution using a single opacity which is likely to be representative of those regions of the disc which control the evolution, taking $\kappa = 0.02 T^{0.8}$ cm$^2$ g$^{-1}$, independent of density $\rho$ \citep{Bell1994, Bell1997}. The initial disc was taken to be an annulus at radius $R_0 = 10$ AU, as described above. The solutions in the $t_\ast - t_{\rm disc}$ plane are shown in Figure~\ref{fig:alphavary}, for three different values of $\alpha =   0.1, 0.01, 0.001$. A mean value of $\alpha = 0.01$ is known to give disc evolutionary timescales in line with known properties of discs \citep{Hartmann1998}. 

As can be seen from the figure, at late times the discs all tend to the same similarity solution, which in this case gives $t_{\rm disc}/t_\ast \rightarrow 1/\sigma = 4.44$. As before, the loci all approach the eventual power law locus from above, because at early times the disc has yet to extend to the centre, and so the accretion rate is lower than that predicted by the similarity solution. The approach occurs later for smaller values of $\alpha$. 

\subsubsection{Full Opacity}

\begin{figure}
\centering
\includegraphics[width=1\columnwidth]{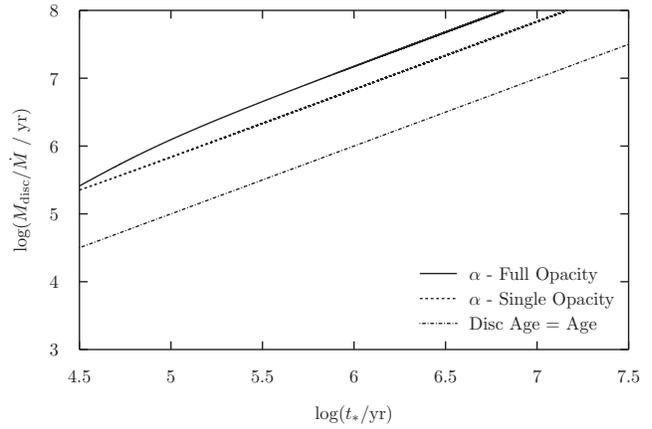}
\caption{Introducing a more physical opacity has little overall effect other than to change the similarity solution the model tends to at late times.  This raising of the disc age means that this, the most physical of the basic models, cannot explain the majority of the observation, which largely fall below the curve.}
\label{fig:multiopc}
\end{figure}

We next computed the evolution of such a disc using a set of opacities given in Table~2. Also given in Table~2 are the values of $1/\sigma$ corresponding to each temperature range (those opacities which give rise to discs which are thermally/viscously unstable are marked with an asterisk in this column.).  In Figure~\ref{fig:multiopc} we compare the loci of a disc using the full opacity table with the disc with only a single opacity. We have chosen one of the larger values of $\alpha$, 0.01, to illustrate the difference in the similarity solution part of the loci. Even for the full opacity, the locus tends to a constant value of $t_{\rm disc}/t_\ast \sim 10$,  because at late times the disc evolution is being controlled by a single low-temperature regime in the disc at large radius. 

\subsubsection{Full Opacity with Delayed Infall}

As we have seen, although the $t_{\rm disc} = t_\ast$ line provides an adequate fit to the data, the standard theoretical disc models with realistic physical opacities tend to produce loci in the $t_\ast - t_{\rm disc}$ plane which lie factors of 4 -- 10 above it . Altering the evolutionary timescale of the discs, by changing the initial disc radii or the size of the viscosity merely exacerbates the problem. 

Rather than start with all the mass in the disc at time $t=0$ we have therefore experimented with the more realistic assumption that the mass is added to the disc on some finite timescale. Usual estimates for the timescale of disc formation are in the region of $10^5$ years \citep{Larson1969,Lin1990}. 

\begin{figure}
\centering
\includegraphics[width=1\columnwidth]{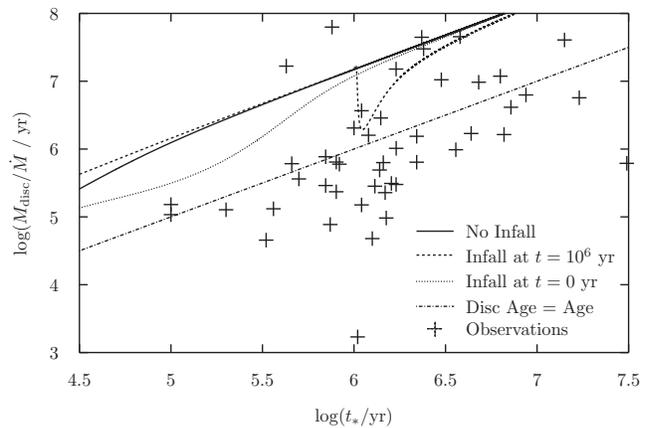}
\caption{Comparison of three alpha prescription discs (with full opacity), the first has no mass infall, the second is an initially low mass ($0.02\;M_{\odot}$) disc with $0.01\;M_{\odot}$ of mass infall around $t=1\times10^{6}\;\mathrm{yr}$, and the third begins with mass infall at $3\times10^{-6}\;M_{\odot}\,\mathrm{yr}^{-1}$ with a fall off time of $10^{5}\;\mathrm{yr}$.  The mass infall has a rejuvenation effect on the disc age, but even doubling the mass of a disc in a small fraction of its lifetime does not produce a large enough effect to to drop below the line $t_{\rm disc} = t_\ast$. Also plotted are the observational data.  The majority of the observations are out of reach with these models alone, as under reasonable constraints they cannot be made to drop below the line $t_{\rm disc} = t_\ast$.}
\label{fig:obscomp}
\end{figure}

In Figure~\ref{fig:obscomp} we show two additional examples of the resulting behaviour. For the first, we start at $t=0$ with a disc annulus of mass $M = 0.02 M_\odot$, and then allow an additional infall of $0.01 M_\odot$ onto the annulus to start at a time of $t = 10^6$ yr. The initial accretion rate at that stage is taken to be $ 1 \times 10^{-7} M_\odot$ yr$^{-1}$ and decays exponentially with a timescale of $10^5$ yr. For the second, the initial disc annulus of $0.2 M_\odot$ is fed at an initial rate of $3 \times 10^6 M_\odot$ yr$^{-1}$, which falls off exponentially on a timescale of $10^5$ yr. 

As can be seen from Figure~\ref{fig:obscomp}, late or delayed infall does indeed have the effect of rejuvenating the disc, and so bringing the loci in the $t_\ast - t_{\rm disc}$ plane below the locus of the similarity solution. Also shown in Figure~\ref{fig:obscomp} are the observational data points. It is evident that even late or delayed infall does not really have much impact in trying to bring theoretical disc predictions into line with the data.

\subsection{Complications to the basic model}

In the above we have considered a basic model for disc evolution in which the matter is added to the disc in a simple fashion and the disc is allowed to evolve using the standard $\alpha$-prescription for the viscosity. However, in reality disc evolution is more complicated that this. We consider here three additional complications. First we consider the model of a layered disc, in which the standard viscosity mechanism is rendered inoperative if the disc is too cool, so that the magneto-rotational instability cannot give rise to MHD-turbulence. Second, we consider what happens in the later stages of the disc's evolution when the disc is evaporated by radiation from the central star. And, third, we consider the formation of a planet which is massive enough to affect disc evolution. We treat each of these effects in fairly idealised form in order to illustrate the likely magnitude and direction the effect has on the loci in the $t_\ast - t_{\rm disc}$ plane. We consider each in turn.

\subsubsection{Layered Disc - Model}

Proto-stellar discs are thought to have episodic accretion histories \citep{Enoch2009}. Such accretion outbursts have been observed and are known as FU Orionis events \citep{Hartmann1996}.  One possible model for this type of phenomena is that of a layered disc.

The model used here is primarily based on that of \cite{Armitage2001}.  Magnetohydrodynamic (MHD) turbulence is assumed to be the process driving viscosity (at small radii), but to be operative this requires the gas to be sufficiently ionised in order for it to interact with magnetic fields.  If the mid-plane temperature $T_{c}$, in the disc is above a critical value $T_{\rm{crit}} = 800\;\rm{K}$ then the gas is assumed to be MHD 'active'.  The top and bottom surface layers (of surface density $\Sigma_{\rm{layer}} = 100\;\rm{g\:cm^{-2}}$) of the disc are always assumed to be active as they are kept ionised by cosmic rays and/or stellar radiation.  Thus an active surface density is defined by

\begin{equation}
\Sigma_{\rm{a}} = \Sigma\;\;\;\rm{for}\;\;\;T_{\rm{c}}>T_{\rm{crit}}
\end{equation}
\begin{equation}
\Sigma_{\rm{a}} = 2\Sigma_{\rm{layer}}\;\;\;\rm{for}\;\;\;T_{\rm{c}}<T_{\rm{crit}}.
\end{equation}
Assuming the disc is Keplerian ($\Omega = \sqrt{GM_{*}/R^{3}}$), the modified surface density evolution equation becomes

\begin{equation}
\frac{\partial \Sigma}{\partial t} = \frac{3}{R} \frac{\partial}{\partial R} \left[ R^{1/2} \frac{\partial}{\partial R} \left( \nu \Sigma_{a} R^{1/2} \right) \right] + \dot{\Sigma},
\end{equation}
where $\dot{\Sigma}$ represents increase in surface density from infalling material.

Thermal equilibrium and the simplistic vertical structure used in \textsection \ref{sec:basicmod} are assumed as before.  Finally the complete description of the viscosity is two-fold.  The regions active to MHD turbulence are described using the $\alpha$-prescription described in \textsection \ref{sec:basicmod}, but in addition the model also takes into account the effects of self-gravity using a prescription dependent on the Toomre $Q$ parameter \citep{Toomre1964},

\begin{equation}
Q = \frac{c_{\rm{s}}\Omega}{\pi G \Sigma}.
\end{equation}
When $Q$ drops below some critical value, taken here to be $Q_{\rm{crit}}=2$, the effects of gravitational instability are deemed to be important and are accounted for by an additional effective $\alpha_{\rm{grav}}$ parameter in the $\alpha$-viscosity prescription \citep{Lin1987, Armitage2001}.

\begin{equation}
\alpha_{\rm{grav}} = 0.01 \left( \frac{Q_{\rm{crit}}^{2}}{Q^{2}} - 1 \right)\;\;\;\rm{for}\;\;\;Q<Q_{\rm{crit}}
\end{equation}
\begin{equation}
\alpha_{\rm{grav}} = 0\;\;\;\rm{otherwise}.
\end{equation}
This complete model results in a disc structure where material at small radii ($< 0.1 \rm{AU}$) is all highly ionised and MHD active, beyond this region out to a few AU only the surface layers are MHD active and there is a 'dead zone' in the disc mid-plane where material builds up.  Finally, in the cool outer part of the disc gravitational instability is the principle mechanism for angular momentum transport.

\subsubsection{Layered Disc - Results}

\begin{figure}
\centering
\includegraphics[width=1\columnwidth]{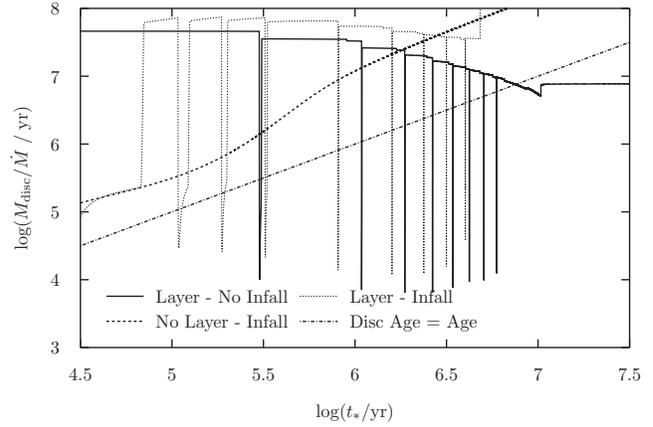}
\caption{Comparison of unlayered and layered discs with and without infalling mass.  The layered discs are clearly different to the unlayered disc. They spend most of their lifetime at high disc age, with intermittent accretion outbursts dropping them briefly to low disc age.  In this way they cover almost the complete range of observed disc ages, but are unlikely to be able to return the same distribution.}
\label{fig:laydisc}
\end{figure}

Introducing a layered disc structure has a dramatic effect on the evolution of the disc.  Figure \ref{fig:laydisc} compares layered and non-layered $\alpha$-prescription models, demonstrating that the forms of disc ages have almost no resemblance to each other.  The infall models have material added to the outer disc (at 10 AU) at an initial rate of $3\times10^{-6}\;\mathrm{M_{\odot}\,yr^{-1}}$, with a fall off time of $10^{5}\;\mathrm{yr}$.  The layered disc with infalling material initially has longer and more frequent accretion outbursts as the dead zone is replenished with mass more rapidly than in the model without infall.  The difference becomes less pronounced at late times (when the infall has decayed away), and is mainly noticeable through the higher disc mass, resulting in higher disc age.

This demonstrates that infalling mass actually has an aging effect on the disc, that is, it makes its disc age greater.  This is somewhat counter-intuitive as one would expect higher mass discs to be younger and less evolved.  The reason for this difference is that unlike in unlayered models the total mass of the disc does not play a significant role in determining the stellar accretion rate during the quiescent periods.  The layered structure means much of the extra mass is held up in the dead zone and does not contribute to the accretion rate (at that time).  Thus, the disc age increases because the mass increases but the accretion rate does not.

Outbursts are triggered when the dead zone becomes so dense that it is gravitationally unstable ($Q<2$).  It then becomes viscous due to gravitational instability, which releases energy, quickly heating the material in the dead zone above the critical temperature for it to be MHD active and then the whole zone rapidly accretes onto the star.  These outbursts drastically alter the disc age of the system and are the only model which enters the region below the disc age line.

The loci of these models in the $t_\ast - t_{\rm disc}$ plane cover almost the entire range of the observed disc ages, but because they spend most of the time in quiescence, i.e. at low accretion rates, in a time-averaged sense the loci are too high to be compatible with the data.  In order to match the data there would need to be a very significant over-detection of outbursting systems, because most of the systems are observed at low disc ages, not at high disc ages, where a layered disc would spend most of its lifetime.  The systems undergoing outbursts may not be in that state for a long period, but they will be far brighter in the UV and hence easier to detect.  Thus, it may be the case that the sample displayed is not representative of the true distribution.  If this were true one would expect the observed distribution of accretion rates to be bimodal, with a significant number of the systems having very high accretion rates ($>10^{-6}\;\mathrm{M_{\odot}\,yr^{-1}}$) and the rest grouped around $10^{-8}\;\mathrm{M_{\odot}\,yr^{-1}}$.  This bimodality should also be reflected in the accretion rates of systems above and below the line $t_{\rm disc} = t_\ast$ if over detection of outbursts were to explain why so many systems have low disc ages. However, inspection of the data indicates that this is not the case.

The layered disc model does however offer a possible explanation for systems observed far above the disc age line.  Many of the disc ages higher than $10^{7}$ years in the observational data are that high, in part, due to the low accretion rates in those systems.  A layered disc spends most of its lifetime in this state, where mass is caught up in a dead zone and the system has a low quiescent accretion rate, resulting in a high disc age.

\subsubsection{Photoevaporation - Model}

Accretion discs around young stars must undergo some form of dispersal as they are only observed to persist for the order of 10 Myr.  A favoured model for this dispersal is photoevaporation, where ionising radiation from the central star causes material in the outer disc to be blown off in a wind.  The model used here is based on that of \cite{Clarke2001}.

Material on the surface of the disc is heated by FUV radiation from the central star. This raises the temperature of the outer disc atmosphere and beyond a certain radius can give it sufficient energy to escape.  This radius is know as the gravitational radius \citep{Hollenbach1994}, and is given by

\begin{equation}
R_{\rm{g}} = \frac{GM_{*}}{c_{\rm{si}}^{2}},
\label{eqn:gravrad}
\end{equation}
where $c_{\rm{si}}$ is the sound speed in the photoionised gas.  Outside of this radius the surface density loss rate due to photoevaporation is

\begin{equation} 
\dot{\Sigma}_{\rm{evap}} = 2 c_{\rm{si}} n_{0}(R) m_{\rm{H}},
\end{equation}
where $m_{\rm{H}}$ is the mass of hydrogen and $n_{0}(R)$ is the number density at the base of the outflow, which is parameterised by

\begin{equation}
n_{0}(r) = n_{0}(R_{\rm{g}})\left(\frac{R}{R_{\rm{g}}}\right)^{-5/2},
\end{equation}
where

\begin{equation}
n_{0}(R_{\rm{g}}) = 5.7\times 10^{4}\,\left(\frac{\Phi}{10^{41}\:\rm{s}^{-1}}\right)^{1/2} \left(\frac{R_{\rm{g}}}{10^{14}\:\rm{cm}}\right)^{-3/2}\;\rm{cm}^{-3}
\end{equation}
and $\Phi$ is the flux of ionising photons emitted by the central star.

Although there are now more sophisticated photoevaporation models (e.g. \cite{Owen2011} and \cite{Alexander2007}) and there is debate about whether FUV, EUV or X-rays play the most important role, all the models exhibit essentially the same phenomenon, with a gap opening in the disc at late times and the inner disc being rapidly accreted.  As here we are principally concerned with ascertaining the effect of such phenomenon on the disc age of a system this model is sufficient for our current purposes. 

\subsubsection{Photoevaporation - Results}

\begin{figure}
\centering
\includegraphics[width=1\columnwidth]{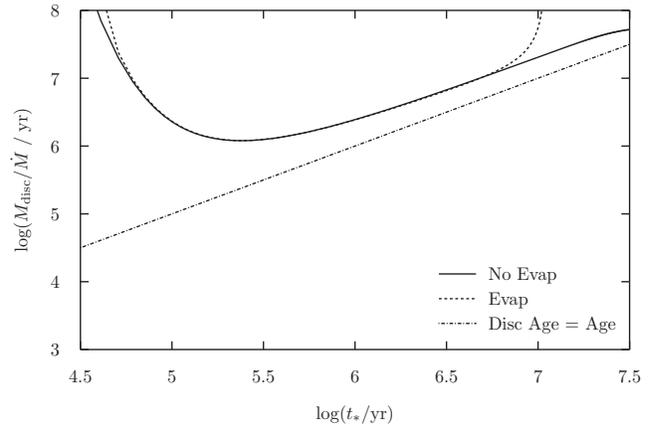}
\caption{The effect of photoevaporation on the disc age of a power law viscosity disc, with initial disc mass of $0.02\;M_{\odot}$.  The only noticeable effect occurs rapidly at late times when the photoevaporation opens a gap in the disc and the inner disc rapidly accretes, resulting in the accretion rate dropping to zero while there is still a finite mass in the outer disc, hence the disc age tends to infinity.}
\label{fig:evap}
\end{figure}

Photoevaporation from the disc has little effect for most of its lifetime, but at late times when the evaporation rate becomes comparable to the accretion rate through the disc, it causes a dramatic change.  A gap in the disc opens up around the gravitational radius (Equation \ref{eqn:gravrad}), this prevents the inner disc from being replenished by material from the outer disc, where most of the mass resides.  The inner disc is rapidly accreted, after which accretion stops and the outer disc is left to gradually evaporate.  Thus at late times the accretion rate drops to zero and the disc age tends to infinity (Figure \ref{fig:evap}).  Therefore, although this may be an important process allowing disc dispersal, it does not help to explain any spread in the disc age.  Photoevaporation has almost no detectable effect on the disc age for the majority of the disc's lifetime, and the period where the disc age increases significantly is brief.  

This is a simplified model and the form of the evaporation would in fact change once the inner disc is removed, but by this time the disc age will already be large, and hence consideration of a more complex photoevaporation model is unlikely to illuminate further effects on the disc age.  However, these 'old' discs may still undergo small outbursts of accretion \citep{Murphy2010}, on the timescale of months down to days.  It is possible that observations of the disc age of such systems could change significantly from survey to survey, but this would of only explain scatter on the upper extreme of the stellar age range observed, leaving the majority of the observations still in need of an explanation.

\subsubsection{Planet Formation \& Migration}

It is widely believed planets  form in the discs around young stars. Furthermore, since many Jupiter mass planets have been observed at small radii where it would not have been possible for them to form, massive planets are also expected to migrate inwards from their formation radius by interaction with the disc.  To model this, we use a simple formulation (based on \cite{Alexander2009}), which considers the effect of planets of around a Jupiter mass on the material in the disc, as the planet accretes material and migrates.

Massive planets exert a torque on material in their vicinity and thus alter the surface density evolution equation.  The evolution due to both planet and disc is described by \citep{Lin1986}

\begin{equation}
\frac{\partial \Sigma}{\partial t} = \frac{1}{R} \frac{\partial}{\partial r} \left[ 3R^{1/2} \frac{\partial}{\partial R} \left( \nu \Sigma R^{1/2} \right) - \frac{2 \Lambda \Sigma R^{3/2}}{\sqrt{GM_{*}}} \right] - \dot{\Sigma}_{\rm{evap}},
\end{equation}
where $\dot{\Sigma}_{\rm{evap}}$ is the photoevaporation as described in the previous section and $\Lambda$ is the torque exerted by the planet, given by

\begin{equation}
\Lambda = -\frac{q^{2}GM_{*}}{2R} \left(\frac{R}{\Delta_{\rm{p}}}\right)^{4}\;\;\;\mathrm{for}\;\;\;R<a
\end{equation}
\begin{equation}
\Lambda = \frac{q^{2}GM_{*}}{2R} \left(\frac{a}{\Delta_{\rm{p}}}\right)^{4}\;\;\;\mathrm{for}\;\;\;R>a,
\end{equation}
where $q=M_{\rm{p}}/M_{*}$ ($M_{\rm{p}}$ is the planet mass), $a$ is the semi-major axis of the planet's orbit and $\Delta_{\rm{p}} = \mathrm{max}(H,|r-a|)$, where $H$ is the scale height of the disc.

In this process angular momentum is transferred between the planet and the material in the disc, this causes the planet to migrate at a rate given by

\begin{equation}
\frac{da}{dt} = -\sqrt{\frac{a}{GM_{*}}} \left(\frac{4\pi}{M_{\rm{p}}}\right) \int_{\rm disc} \! R\Lambda \Sigma \,dR.
\end{equation}
The process of angular momentum exchange also acts as an effective repulsive force away from the planet because angular momentum is removed from material inside the planet's orbit and added to material outside.  Thus, the planet  opens up a gap in the disc.  In this model material is allowed to flow across the gap, some of which the planet will accrete.  The efficiency of this process, $\epsilon$, is approximated by the fitting function

\begin{equation}
\frac{\epsilon (M_{\rm{p}})}{\epsilon_{\rm{max}}} = 1.67\:\left(\frac{M_{\rm{p}}}{M_{\rm{Jup}}}\right)^{1/3} \exp \left(\frac{-M_{\rm{p}}}{1.5\:M_{\rm{Jup}}}\right) + 0.04,
\end{equation}
where $\epsilon_{\rm{max}}$ is taken to be 0.5 \citep{Veras2004, DAngelo2002, Lubow1999}.  Giving the accretion rate across the gap as

\begin{equation}
\dot{M}_{\rm{gap}} = \frac{1}{1+\epsilon} \dot{M}_{\rm{p}}
\end{equation}
and the accretion rate on to the planet as

\begin{equation}
\dot{M}_{\rm{p}} = \epsilon \dot{M}_{\rm{disc}},
\end{equation}
where $\dot{M}_{\rm{disc}}$ is the accretion rate through the disc well outside the planets orbit.  This is calculated by assuming a steady state model at 3 times the planet's orbital radius, giving

\begin{equation}
\dot{M}_{\rm{disc}} = 3\pi\nu(3a)\Sigma(3a).
\end{equation}

\subsubsection{Planet Formation \& Migration - Results}

\begin{figure}
\centering
\includegraphics[width=1\columnwidth]{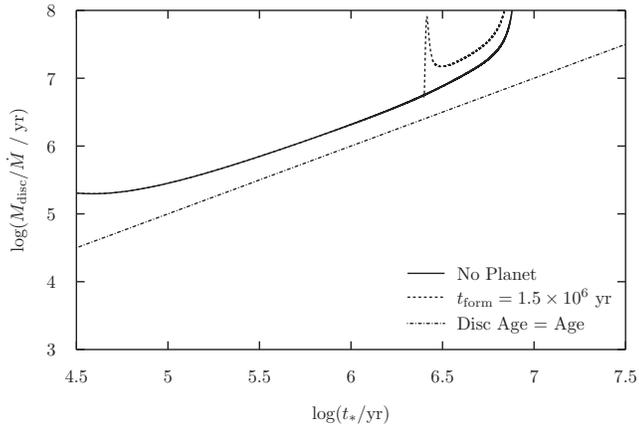}
\caption{Formation of a $0.5\;M_{\mathrm{Jup}}$ planet at $t = 1.5\times10^{6}\;\mathrm{yr}$ in a power law viscosity disc with photoevaporation.  The initial disc mass was $0.03\;M_{\odot}$ to ensure the planet had a noticeable effect on the disc but did not simply fall into the star.  The planet effectively 'pins back' the material at larger radii than its orbit, this slows the accretion rate and hence raises the disc age.}
\label{fig:planet}
\end{figure}

When a massive planet forms it holds back material in the outer disc and prevents it from flowing to the inner disc as rapidly as it otherwise would.  This in turn lowers the accretion rate onto the star, and thus increases the disc age.  After it has formed, the planet accretes mass and migrates inwards.  In the case we compute(Figure \ref{fig:planet}), the final planet mass was $M_p \approx  0.78\;M_{\mathrm{Jup}}$ (initially $0.5\;M_{\mathrm{Jup}}$) and it migrated inwards ending up at $\sim 5.2\;\mathrm{AU}$ from $5.6\;\mathrm{AU}$.  After a period of rapid change immediately post formation the disc age settles down to another, higher disc age, similarity solution which is eventually truncated by photoevaporation (described in previous subsection).
This illustrates the basic effect the formation of a massive planet on the disc age.
But the effect of the planet is always be to increase the disc age by slowing the accretion rate.  Thus the effect of planet formation is to push the disc age of systems to lie further above the line $t_{\rm disc} = t_\ast$.

\section{Discussion}
\label{sec:discussion}

\begin{figure*}
\begin{minipage}{175mm}
\centering
\includegraphics[width=1\columnwidth]{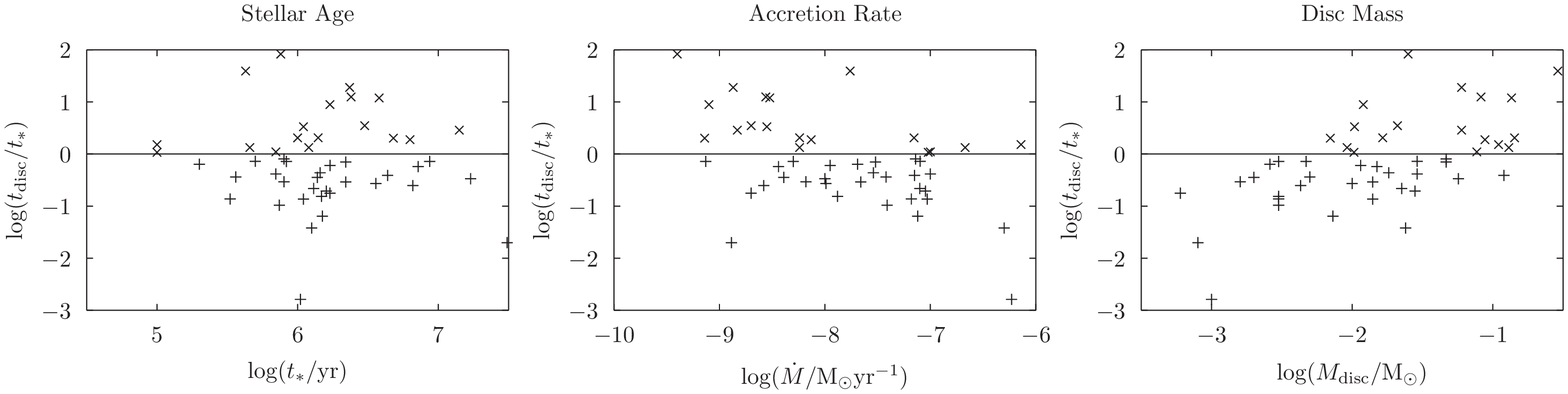}
\caption{Distributions of stellar age, accretion rate and disc mass of the observations compared between those systems which fall above and below the line $t_{\rm disc} = t_\ast$.  The corresponding KS-test P-values are 0.954, 0.049 and 0.003 respectively.  Both the stellar age and accretion rate distributions seem similar for high and low disc ages, but the are significantly more low mass systems at low disc ages, and more high mass systems at high disc ages.  This may indicate there is a systematic error involved in the disc mass measurements. }
\label{fig:distplot}
\end{minipage}
\end{figure*}

\begin{figure}
\centering
\includegraphics[width=1\columnwidth]{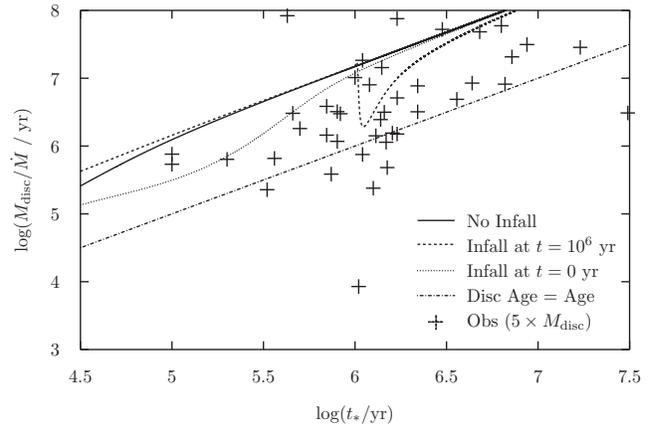}
\caption{Compares the most physical basic models from \textsection \ref{sec:modres} with where the observations would lie if all the disc masses were increased by a factor of 5.  This raises many of the observations into the region above the line $t_{\rm disc} = t_\ast$, where they can potentially be explained by one of the models studied, within the observational scatter. }
\label{fig:obscomp2}
\end{figure}

We have considered the predicted evolution of theoretical accretion discs in the $t_\ast - t_{\rm disc}$ plane, where 
\begin{equation}
t_{\rm disc} = \frac{M_{\rm disc}}{\dot{M}},
\end{equation}
and have compared the results with observational data, using the estimated age of the central star, $t_\ast$, as a proxy for the age of the disc, $t$. We noted that, despite large scatter, and large uncertainties in the observational estimates, the observational data are approximately fit by the relationship $t_{\rm disc} = t_\ast$ (Figure~\ref{fig:obsplot}) with a scatter of around an order of magnitude above and below this line.

We showed that if the discs are modelled using viscosity prescriptions which are either a simple power law or the standard alpha model, then the loci in the $t_\ast - t_{\rm disc}$ plane quickly evolve to a line of the form $t_{\rm disc}/t_\ast = 1/\sigma$. For the power law of the form $\nu \propto R$, we find $1/\sigma = 2$ \citep{Hartmann1998}, but for more physically realistic accretion disc models, using the $\alpha$-prescription, and local disc structure computations, the values of $1/\sigma$ lie more in the range 3 -- 10 (Figure~\ref{fig:multiopc}). 

Thus the simple theoretical disc models appear to display disc ages which are too high by factors of order 3 -- 10. This remains true, even when late or delayed infall onto the discs is taken into account.

Given this, we have investigated three further physical effects which can have a bearing on the measured disc age. These are
\begin{enumerate}
\item It may be that those parts of the disc which are sufficiently cool that the MRI cannot operate, and so cannot provide a viscosity.  In this case, accretion initially occurs as a series of outbursts \citep{Armitage2001}. This has a profound impact on the locus the disc follows in the $t_\ast - t_{\rm disc}$ plane.  The locus in this case is capable of covering the area in the plane occupied by observed systems (Figure~\ref{fig:laydisc}). However, in these models the discs spend most of their time at very large values of $t_{\rm disc}$, with only brief sorties to much lower values (during the outbursts, when the accretion rates are high). This model could therefore account for the observations, but there would need to be very strong selection effects against the detection of systems with low accretion rates.
\item In the standard models all the disc material is assumed to end up accreting onto the central star. In reality, it is possible that at late stages the disc material is blown away by the central star. 
The effect of this is to reduce the accretion rate onto the star, at late times, and so to correspondingly increase the disc age still further. 
\item One of the expectations of these 'proto-planetary' discs is that at least some of them succeed in forming planets. Formation of a planet in the disc which is large enough to interrupt disc accretion, and to allow migration of the planetary orbit, also has the effect of decreasing the central accretion rate, and so of increasing the disc age.
\end{enumerate}
It is therefore difficult for current models of the evolution of proto-planetary discs to provide a satisfactory fit to the observed points, taken at face value, in the $t_\ast - t_{\rm disc}$ plane.  Hence it is worth enquiring whether some systematic effect or effects might lead one or more of the observational quantities to be mis-estimated. The three relevant quantities are (i) stellar age, (ii) disc mass, and (iii) accretion rate. 

If one of these quantities was primarily responsible for the observed disc ages to appear artificially low, it might be that its estimated value varies systematically with disc age. Thus we have investigated separately how each of the distributions of stellar age, accretion rate and disc mass compare, between those data points that lie  above and below the $t_{\rm disc} = t_\ast$ line. From Figure \ref{fig:distplot} it is evident the distributions of stellar age and of accretion rate for those objects with high and low ratios of $t_{\rm disc}/t_\ast$ are not significantly different.  This is confirmed by the 2-sided K-S test on the distributions for high and low disc age relative to stellar age (points either above or below the line), which returns probabilities that these are drawn from the same distribution of 0.954 and 0.049, respectively. However, the disc mass distribution does seem to show significant dependence on the value of $t_{\rm disc}/t_\ast$, and the K-S test indicates the probability of only 0.003 that the disc masses of objects with $t_{\rm disc} > t_\ast$ and $t_{\rm disc} < t_\ast$ are drawn from the same distribution (shown graphically in Figure \ref{fig:distplot}).  Of course because disc age is an estimator of stellar age and is dependent on both $M_{\rm disc}$ (linearly) and $\dot{M}$ (inversely) all of these quantities are interdependent.  This clouds the meaning of the plots in Figure \ref{fig:distplot} somewhat, but what is clear is that the value of $M_{\rm disc}$ is dependent on whether the system has a high or low disc age relative to its stellar age, where as the values of $\dot{M}$ and $t_\ast$ are not.  In particular, objects with low values of $t_{\rm disc}/t_\ast$ tend to have on average lower values of $M_{\rm disc}$.  

To explain the discrepancy through either the stellar age or the accretion rate, one or both of these quantities would need to be systematically overestimated.  Although the K-S tests hint towards the disc masses being the culprit of the discrepancy we will discuss the potential for mis-estimation of the other variables in the disc age.

Measurement of accretion rates contains several potential sources of systematic errors \citep{Calvet2000, Calvet2011}.  To measure the UV excess the emission from the star must be deducted, but stellar models do not always fit well to young stars.  A bolometric correction must be applied to calculate the accretion luminosity, the error here could be up to a factor of 2 if line emission dominates over the continuum.  The conversion between flux and luminosity also requires the distance to the star to be known.  Finally, converting to the accretion rate requires knowledge of the mass and radius of the star and an assumption regarding inner disc geometry.  These errors are both significant and inter-dependent, and it is not clear if their combined effect will be in any particular direction or have an overall random effect on a large sample.

Ages of young stars are measured by fitting theoretical pre-main-sequence evolutionary tracks to data in the H-R diagram, the position of which is derived by determining the stellar luminosity and its spectral type.  The process is clouded by the presence of an accretion disc which introduces complications in determining the correct luminosity (see \textsection \ref{sec:age}), this uncertainty is then increased as there is no consensus as to which theoretical track is the best fit, hence different authors derive different age estimates, even for the same initial data.  \cite{Hillenbrand2009} concluded that this method likely under-predicts low mass stellar ages by 30 -- 100\% and \cite{Naylor2009} recently developed a new fitting procedure, deriving stellar ages of young clusters that were a factor of 1.5 -- 2 longer than those commonly found, suggesting that current age estimates may be systematically too short.  Thus, the errors in stellar ages are likely in the opposite direction to that which would assist in explaining the discrepancy we have found.

We therefore speculate that there might be a systematic error in the measurement of disc masses. We note that if {\em all} disc masses were currently being systematically underestimated by a factor of 3 -- 5, then we would achieve approximate agreement between the predictions of theoretical models and the observations. We show this in Figure~\ref{fig:obscomp2}.

The suggestion that there is a tendency to underestimate the masses of proto-planetary discs is not a new one, \cite{Hartmann2006} suggests that grain growth and the resulting reduction in dust opacity could lead to a systematic underestimation of disc masses. The basic problem is that disc mass is estimated from measurement of far-IR emission from dust grains.  There may be a tendency for the gas to be evaporated, or otherwise lost, from discs at a greater rate than the grains. This would lead to disc masses being overestimated.  Another problem is that measurement of disc mass relies on assumptions about the relative number of smaller grains (which dominate the radiative emission) and the larger grains (which dominate the grain mass). If grains tend to coagulate with time (a process which is a necessary precursor to the formation of planets) then the ratio of larger to smaller grains grows with time, and disc masses would be underestimated.  \cite{Williams2011} suggest that the latter dominates, and that the overall tendency is that dust masses are underestimated.

\section*{Acknowledgements}
RDA acknowledges support from the Science \& Technology Facilities Council (STFC) through an Advanced Fellowship (ST/G00711x/1).

\bibliographystyle{mn2e}

\bibliography{jones_pringle_alexander}

\end{document}